\documentclass[journal=jacsat,manuscript=article]{achemso}

\usepackage{chemformula} % Formula subscripts using \ch{}
\usepackage[T1]{fontenc} % Use modern font encodings

\author{Himanshu Bangar}
\email{himhim@chalmers.se}
\affiliation{Department of Microtechnology and Nanoscience, Chalmers University of Technology, SE-41296, Göteborg, Sweden}

\author{Polychronis Tsipas}
\affiliation{National Center for Scientific Research DEMOKRITOS 15341, Athens, Greece}

\author{Prasanna Rout}
\affiliation{Department of Microtechnology and Nanoscience, Chalmers University of Technology, SE-41296, Göteborg, Sweden}
\author{Lalit Pandey}
\affiliation{Department of Microtechnology and Nanoscience, Chalmers University of Technology, SE-41296, Göteborg, Sweden}
\altaffiliation{Wallenberg Initiative Materials Science for Sustainability, Department of Microtechnology and Nanoscience, Chalmers University of Technology, SE-41296, Göteborg, Sweden}

\author{Alexei Kalaboukhov}
\affiliation{Department of Microtechnology and Nanoscience, Chalmers University of Technology, SE-41296, Göteborg, Sweden}

\author{Akylas Lintzeris}
\affiliation{National Center for Scientific Research DEMOKRITOS 15341, Athens, Greece}
\altaffiliation{School of Applied Mathematical and Physical Sciences, National Technical University of Athens, 157 80 Athens, Greece}

\author{Athanasios Dimoulas}
\email{a.dimoulas@inn.demokritos.gr}
\affiliation{National Center for Scientific Research DEMOKRITOS 15341, Athens, Greece}

\author{Saroj P. Dash}
\email{saroj.dash@chalmers.se}
\affiliation{Department of Microtechnology and Nanoscience, Chalmers University of Technology, SE-41296, Göteborg, Sweden}
\altaffiliation{Wallenberg Initiative Materials Science for Sustainability, Department of Microtechnology and Nanoscience, Chalmers University of Technology, SE-41296, Göteborg, Sweden}
\alsoaffiliation{Graphene Center, Chalmers University of Technology, SE-41296, Göteborg, Sweden}

\title{Interplay between altermagnetic order and crystal symmetry probed using magnetotransport in epitaxial altermagnet MnTe}

\abbreviations{IR,NMR,UV}


\begin{document}

\begin{abstract}

Altermagnets are a new class of magnetic materials characterized by fully compensated spins arranged in alternating local structures, allowing for spin-split bands similar to those found in ferromagnets without net magnetism. Recently, MnTe has emerged as a prototypical altermagnetic material exhibiting spin-polarized electronic bands and anomalous transport phenomena. Although recent work has explored the magnetic and structural properties of MnTe, detailed experimental investigations into the relationship between altermagnetic order and crystal symmetry are lacking. Here, we report the relationship between altermagnetic order and crystal symmetry by investigating magnetotransport properties of MnTe epitaxial altermagnetic thin films grown by molecular beam epitaxy. %We perform systematic angle-dependent measurements to probe the interplay between altermagnetic order and crystal symmetry. 
We observe a spontaneous anomalous Hall effect and show the control of Hall response with the altermagnetic order using the magnetic field and the crystallographic angle dependence. Detailed measurements establish that both the longitudinal and transverse electronic responses depend on the relative orientation of the applied current and
Néel vector as well as on the crystal orientation and altermagnetic order. These results provide new insights into the interplay between crystal symmetry and altermagnetism for future device applications.
 
\end{abstract}

\section{Introduction}
\begin{figure}[t!]
\centering
\includegraphics*[width=\textwidth]{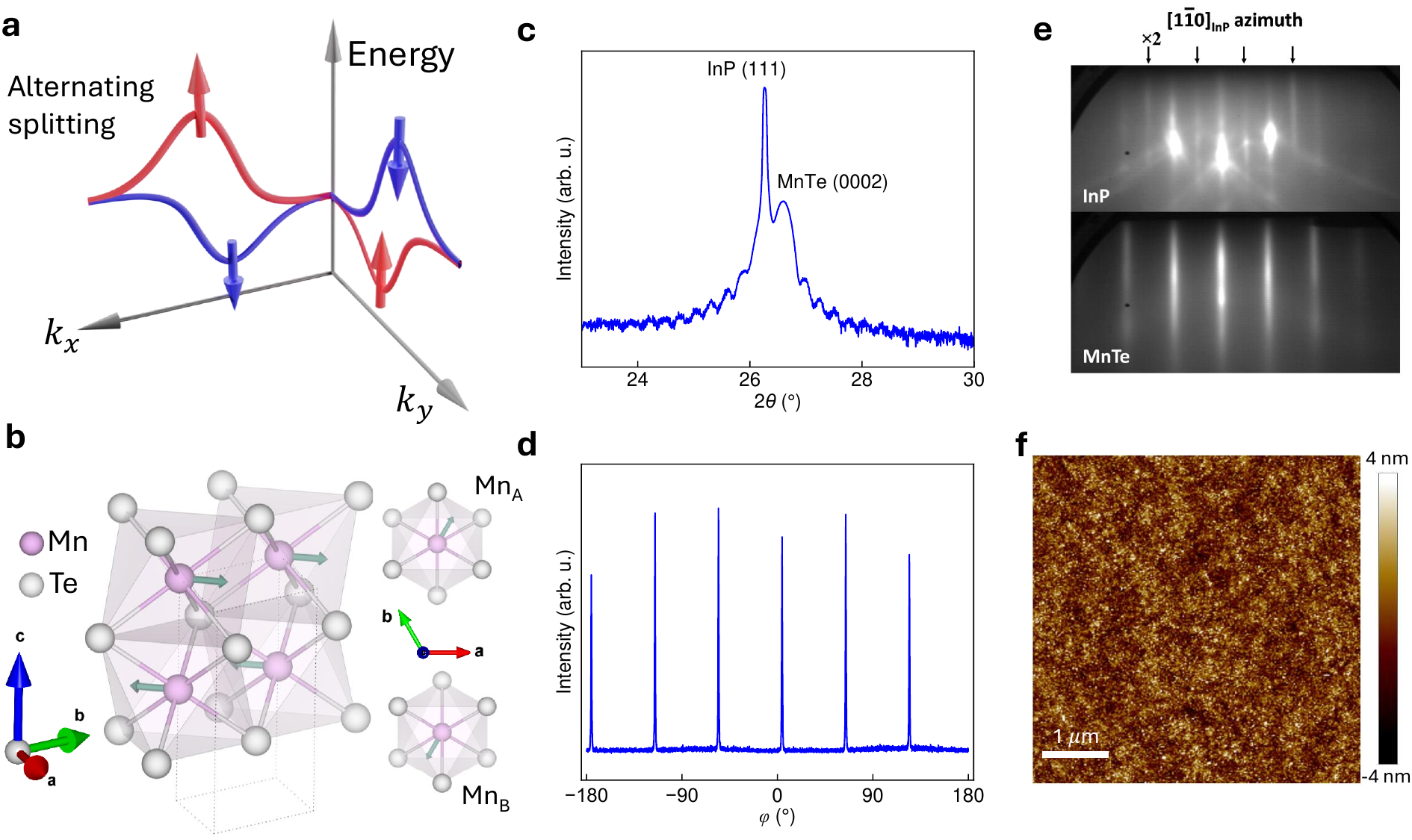}
\caption{\label{fig:1} \textbf{ Altermagnet MnTe epitaxial thin film characterization} \textbf{(a)} Anisotropic splitting of Fermi surfaces is a fingerprint of altermagnets. \textbf{(b)} Crystal structure of the $\alpha$-MnTe. \textbf{(c)} X-ray diffraction (2$\theta-\omega$) scan showing MnTe (0002) peak along with InP (111) peak. \textbf{(d)} $\varphi$ scan centered on MnTe (10$\bar{1}$2) peak, revealing six fold symmetry associated with the hexagonal crystal structure of MnTe. \textbf{(e)} In situ RHEED patterns of the InP (111) substrate and the MnTe flim. \textbf{(f)}. Atomic force microscopy (AFM) scan of the MnTe (40  nm)/Al$_2$O$_3$ (2 nm) thin film showing a smooth surface.%Scanning tunneling microscope (STM) image revealing the topography of 40 nm thick MnTe film. The scan was performed in-situ immediately after the film deposition.
}
\end{figure}

Altermagnets represent a recently proposed class of magnetically ordered materials that provide exciting avenue for exploring novel physical phenomena due to their unique symmetry properties~\cite{vsmejkal2022beyond,vsmejkal2022emerging,yuan2021prediction}. Unlike ferromagnets, which exhibit a net magnetization, or traditional antiferromagnets, which are globally symmetric, altermagnets exhibit collinear altermagnetic order with zero net magnetization but broken spatial symmetries that enable novel transport responses~\cite{krempasky2024altermagnetic,mazin2023altermagnetism,feng2022anomalous,vsmejkal2020crystal}. A defining characteristic of altermagnets is their unconventional non-relativistic spin-splitting of electronic bands in momentum space as presented in Fig.~\ref{fig:1}a.~\cite{bai2024altermagnetism,guo2023spin} This momentum-dependent spin splitting arises from the unique crystal symmetries such as crystal rotations or glide planes connecting the opposite-spin sublattices, leading to broken Kramers degeneracy even in the absence of spin-orbit coupling.~\cite{lee2024broken,yang2025three,reimers2024direct} These properties can have significant implications as they provide new avenues for manipulating spin currents and developing novel devices based on antiferromagnetic materials, which offer advantages such as faster spin dynamics and less influence to stray magnetic fields.~\cite{song2025altermagnets,cheong2024altermagnetism,sourounis2025efficient,bangar2023large,amin2024nanoscale,li2024topological,hu2024spin,gomonay2024structure}

Manganese telluride (MnTe) has emerged as a prototypical altermagnet drawing huge interest for its unconventional magnetic and transport properties.~\cite{kravchuk2025chiral,amin2024nanoscale,kriegner2016multiple} Structurally, MnTe crystallizes in a hexagonal lattice with a collinear antiferromagnetic order as illustrated in Fig.~\ref{fig:1}b.~\cite{gonzalez2023spontaneous,osumi2024observation}. Its magnetic structure is characterized by alternating magnetic sublattices represented by Mn$_{\rm A}$ and Mn$_{\rm B}$ in Fig.~\ref{fig:1}b, which are related by crystal rotational symmetry instead of lattice translation or spatial inversion symmetry as in conventional antiferromagnets.~\cite{lovesey2023templates,takahashi2025symmetry} These symmetry properties have important implications on the transport properties of MnTe, enabling the observation of anomalous Hall effect (AHE) even in a non-ferromagnetic system.~\cite{gonzalez2023spontaneous,kluczyk2024coexistence,sato2024altermagnetic} The AHE in these materials is attributed to the unique spin splitting of the electronic bands, driven by the crystal symmetry and altermagnetic ordering, leading to a non-zero Berry curvature in momentum space.~\cite{xiao2010berry,vsmejkal2022anomalous} A recent study has suggested that the sign of the anomalous Hall effect in an altermagnet should reverse upon reversal of the Néel vector, and that its very existence depends critically on the orientation of the Néel vector relative to the crystal axes.~\cite{tschirner2023saturation} Such dependence of the AHE on crystal symmetry implies that MnTe should exhibit anisotropic magnetotransport when an external magnetic field and/or current is applied along different directions relative to the crystal axes. While previous studies have examined various magnetic and structural properties of MnTe, detailed investigations into the interplay between Néel vector orientation and crystal symmetry through transport measurements has not been done yet. Addressing this gap is necessary for understanding and advancing the altermagnetic phenomena.

In this report, we demonstrate that the magneto-transport in MnTe is governed by the relative orientation of the crystal symmetry and altermagnetic order. In epitaxial altermagnetic MnTe thin films, we observe a spontaneous anomalous Hall effect and show the control of Hall response in detail for the first time. We found that the longitudinal and transverse resistivity depend on the relative orientation of the current and altermagnetic order, as well as on the crystal orientation and altermagnetic order. Understanding the angle and crystal -dependent magnetotransport  holds implications for controlling altermagnetic Néel vector for future device application.

\section{Results and discussion}

$\alpha$-MnTe crystallizes in the hexagonal NiAs structure (crystallographic space group P63/$mmc$ 194)~\cite{Villars2023:sm_isp_sd_0530446} as depicted in Fig.~\ref{fig:1}b. The magnetic moments of two Mn atoms in the unit cell align antiparallel within the ab-plane (basal plane) of the material with a preferential easy axis alignment along the [01$\bar{1}$0] or the other two equivalent crystal directions. We grow 40 nm thick epitaxial MnTe (0001) thin films using molecular beam epitaxy (MBE) on InP (111) substrate (methods for growth details). The in-plane lattice constant of 0.4151 nm for the InP (111) surface closely matches with a = 0.4149 nm for hexagonal $\alpha$-MnTe.~\cite{Villars2023:sm_isp_sd_0530446,Villars2023:sm_isp_sd_0453882} This minimal lattice mismatch (0.048\%) makes InP a suitable substrate for the epitaxial growth of MnTe, which is investigated using X-ray diffraction (XRD). The 2$\theta-\omega$ scan presented in Fig. 1c shows the expected (002) peak of alpha-MnTe (supplementary Fig.~\ref{fig:S1}a for full scan). In addition, the presence of Laue thickness fringes establishes the highly crystalline and low roughness quality of our films. To investigate the epitaxial nature of our films, we performed $\varphi$ scan centred on MnTe (10$\bar{1}$2) peak as shown in Fig.~\ref{fig:1}d. We observed a clear six-fold symmetry corresponding to the hexagonal basal plane of MnTe film. Moreover, we performed in-situ reflection high-energy electron diffraction (RHEED) patterns of the InP (111) substrate and the MnTe film in Fig.~\ref{fig:1}e. The x2 reconstruction of the InP indicates a clean substrate surface. The alignment of the main streaks between MnTe and InP shows nearly perfect lattice matching because of epitaxial growth. The surface topology is investigated by performing Atomic force microscopy (AFM) scan. Figure~\ref{fig:1}f presents a AFM scan of the MnTe (40  nm)/Al$_2$O$_3$(2 nm) film. The root mean square roughness is determined to be 1.09 nm showing a smooth surface.

%an in situ scanning tunneling (STM) scan. Figure~\ref{fig:1}f presents a typical STM scan of the MnTe film that reveals the formation of 2D islands. The root mean square (rms) roughness over a 500×500 nm$^2$ area is 0.3 nm showing a smooth surface.

\begin{figure}[t!]
\centering
\includegraphics*[width=1\textwidth]{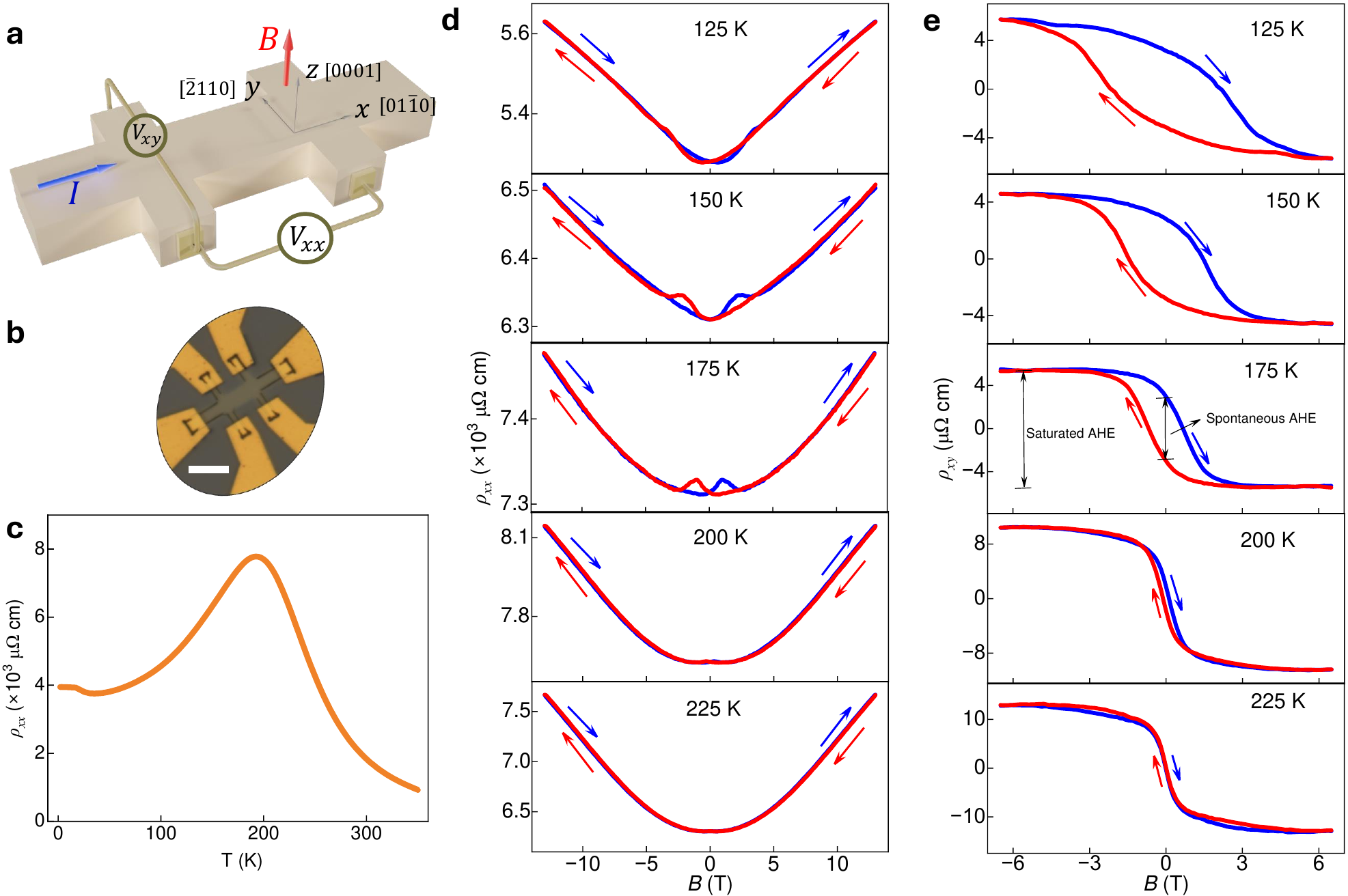}
\caption{\label{fig:2} \textbf{Magnetotransport measurements in Altermagnet MnTe thin film.} \textbf{a} Schematic of the Hall bar device. \textbf{b} The optical image of measured device. Scale bar is 7 µm. \textbf{c} Temperature dependent longitudinal resistivity of 40 nm thick $\alpha$-MnTe film. \textbf{d} Magneto-resistance and \textbf{e} anomalous Hall effect responses at different temperatures above and below the Néel temperature $\approx$ 200 K.} 
\end{figure}

To investigate longitudinal and transverse electronic transport in MnTe films, we fabricated Hall bar devices as shown in Fig.~\ref{fig:2}a,b (see methods for fabrication details). The current $I$ is applied along $x$-axis, which corresponds to [01$\bar{1}$0] crystal direction. We present the temperature dependence of the longitudinal resistivity ($\rho _{xx}$) in Fig.~\ref{fig:2}c, which follows a trend similar to that reported in previous reports~\cite{gonzalez2023spontaneous,magnin2012monte}. The peak observed in $\rho _{xx}$(T) signifies the Néel temperature (T$_{\rm N}$), which is approximately 200 K in our samples. In order to further probe the magnetic nature, we measure $\rho _{xx}$ and  $\rho _{xy}$ at different temperatures by sweeping magnetic field ($B$) in an out-of-plane direction to the sample surface (Fig.~\ref{fig:2}d). Above T$_{\rm N}$, no hysteresis is observed in the magnetoresistance (MR), i. e. $\rho _{xx} (B)$, as expected from the paramagnetic state. We see a clear hysteretic behavior in the MR below T$_{\rm N}$ as MnTe enters the antiferromagnetically ordered spin structure within the hexagonal basal plane. We see a hysteretic behavior due to domain reorientation or a spin-flop transition where the sub-lattices reorient relative to the applied field.~\cite{kriegner2017magnetic,kluczyk2024coexistence,bogdanov2007spin} In addition to MR, we simultaneously measured the Hall signal $\rho _{xy} (B)$, as shown in Fig.~\ref{fig:2}e. The presented data are obtained after subtracting the ordinary Hall effect and in-field contribution by following the process mentioned in supplementary section S2. We observe a clear hysteresis in $\rho _{xy} (B)$ below T$_{\rm N}$, which is attributed to anomalous Hall effect (AHE) arising from alternating spin splitting and the associated Berry curvature.~\cite{gonzalez2023spontaneous} With decreasing temperature, we observe a gradual increase in the coercivity (Fig.~\ref{fig:2}d,e), which clearly highlights the change in magnetic anisotropy with temperature. Moreover, the closing of the hysteresis loop in both MR and AHE signals the spin-flop transition (T$_{\rm sf}$), where the altermagnetic order fully aligns with the applied magnetic field. At 175 K, we determine spin-flop field ($B_{\rm sf}) \approx$ 2.5 T from the magnetotransport measurements presented in Fig.~\ref{fig:2}d,e. Moreover, the behaviour of the saturated and spontaneous AHE signal is presented in Fig.~\ref{fig:S3}. The complex trend observed is possibly originating from a combination of several factors including changes in Fermi level and level of polarization with temperature.~\cite{gonzalez2023spontaneous} 

% A change in anisotropy due to a phase change at low temperatures was previously reported in MnTe.~\cite{hennion2002spin,krause2013structural} which further confirms the altermagnetic nature of MnTe

% ence, further in this manuscript, we perform magnetic field  and crystallographic angular dependence at 175 K where we have the altermagnetic phase.

\begin{figure}[t!]
\centering
\includegraphics*[width=\textwidth]{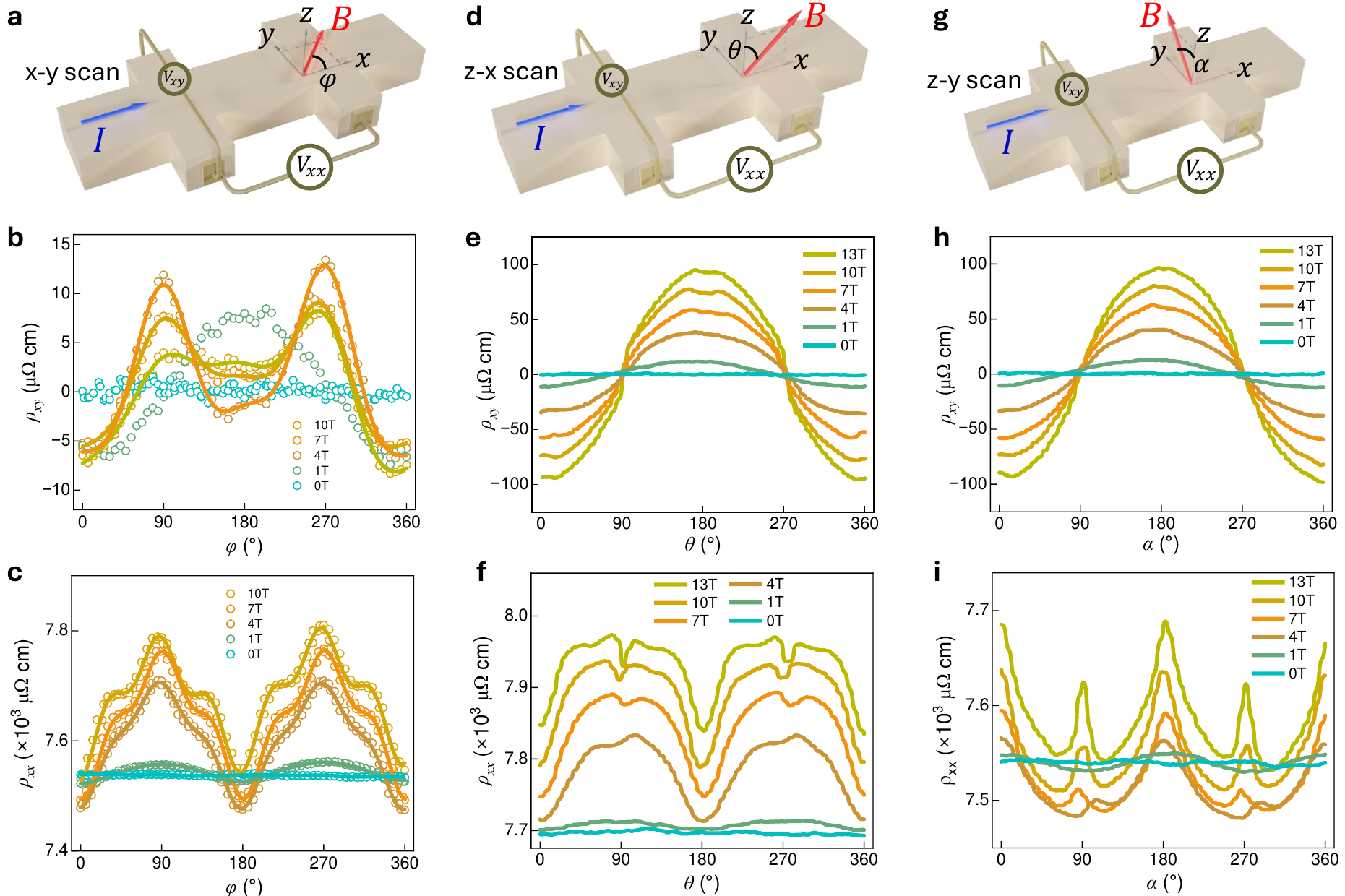}
\caption{\label{fig:3} \textbf{Transverse and longitudinal magnetotransport in three different geometries.} \textbf{a,d,g} display three measurement geometries, in which the magnetic field is rotated in xy, zx, yz planes, respectively. \textbf{b,c} The angular dependence of transverse and longitudinal resistivities (open circle) at different magnetic fields in xy-plane while the solid lines are the fits as discussed in the text. Similar angular dependence of $\rho_{xy}$ and $\rho_{xx}$ in zx-plane are presented in \textbf{e,f} and those in yz-plane are presented in \textbf{h,i}. All the measurements were performed at 175 K (below Néel temperature).} 
\end{figure}

% provide crucial insights into the interplay between the altermagnetic order and the electronic transport properties of MnTe. By varying the orientation of the magnetic field with respect to the crystal axis and the direction of the electric current, one can probe the anisotropy of the AHE and MR.~\cite{gonzalez2024anisotropic} The AHE, in particular, is sensitive to the underlying magnetic structure and the symmetry of the electronic band structure.~\cite{wu2025optical,zhou2025manipulation} These measurements can reveal the symmetry of the spin-dependent scattering processes and the topology of the electronic band structure, offering a deeper understanding of the underlying mechanisms responsible for the AHE in this material. Magnetoresistance can also provide valuable information about altermagnetic ordering and the scattering mechanisms in MnTe.~\cite{kriegner2017magnetic} Therefore, we investigate the angular dependence of the magnetotransport response in-plane and out-of-plane in three different geometries, as shown in Fig.~\ref{fig:3}(a,d,g).

Further insight into the altermagnetic order in MnTe films can be obtained from the angular dependence of magnetotransport, which probes how the electronic response varies with the Néel vector and crystal orientation~\cite{rout2017sixfold,gonzalez2024anisotropic,ritzinger2023anisotropic}. We investigate the angular dependence of transverse and longitudinal resistivity at 175 K in three different measurement geometries, as depicted in Fig.~\ref{fig:3}(a,d,g). First, we present the $\rho_{xy}$ and $\rho_{xx}$ measured for the fields applied in the $xy$ plane (Fig.~\ref{fig:3}b and c, respectively), where we vary the angle $\varphi$ between the current direction along $x$-axis and the magnetic field $B$ as shown schematically in Fig.~\ref{fig:3}a. %Please note that above $B_{\rm sf}$, we can refer $\varphi$ as angle between current and Néel vector.
The angular dependence can be divided into two regimes: low-field regime (B$\leq$ 1 T) and high-field regime (B$\geq$ 4 T). In the low-field regime, the Néel vector does not change (for $B =0$) or fully rotate with the magnetic field (for $B =1$~T). Therefore, we do not observe any substantial angular dependence of $\rho_{xy}$ and $\rho_{xx}$. In contrast, in the high-field regime, the Néel vector undergoes full rotation with the magnetic field, leading to a pronounced angular dependence. This indicates that the magnetization starts aligning as the field increases, and the spin-flop transition occurs between these two regimes.

To better understand the angular dependence, we consider the symmetry analysis for a rotation of the Néel vector within the basal plane, which leads to two types of AMR. The first one is twofold non-crystalline AMR, which depends on the relative orientation of the Néel vector with respect to the current.~\cite{McGuire1975anisotropic,ritzinger2023anisotropic} The other is crystalline AMR with higher-order terms as a result of the crystal symmetry of the lattice.\cite{rout2017sixfold,gonzalez2024anisotropic} In case of MnTe with hexagonal basal plane, $\rho_{xx} = \rho_{2} {\rm cos}(2\varphi_I - 2\varphi) + \rho_{4} {\rm cos}(2\varphi_I + 4\varphi)  + \rho_{6} {\rm cos}(6\varphi)$ and $\rho_{xy} = -\rho_{2} {\rm sin}(2\varphi_I - 2\varphi) - \rho_{4} {\rm sin}(2\varphi_I + 4\varphi)$, where the angle between the $x$-axis and the current direction is $\varphi_I = 0$, while the angle between the $x$-axis  and the Néel vector is $\varphi$, and $\rho_{i}$ is the the amplitude of $i$-th order contribution. In addition, we expect a threefold $\rho_{xy}$ component corresponding to the AHE due to the altermagnetic nature of our system.~\cite{gonzalez2024anisotropic,kluczyk2024coexistence,gonzalez2023spontaneous} We can simplify these expressions in the high-field regime ($B > B_{\rm sf}$), where the Néel vector aligns with the magnetic field direction. In this case, $\varphi$ denotes the angle between the $x$-axis and the magnetic field direction [see Fig.~\ref{fig:3}a]. The expressions for the longitudinal and transverse resistivities then reduce to: $\rho_{xx} = \rho_{2} {\rm cos}(2\varphi) + \rho_{4} {\rm cos}(4\varphi)  + \rho_{6} {\rm cos}(6\varphi)$ and $\rho_{xy} = -\rho_{2} {\rm sin}(2\varphi) + \rho_{3} {\rm sin}(3\varphi) - \rho_{4} {\rm sin}(4\varphi)$. Figures \ref{fig:3}b,c confirm that the measurements at 4, 7, and 10 T are well described by the corresponding expressions. Both $\rho_{xx}$ and $\rho_{xy}$ are dominated by $2\varphi$ component (see SI Fig.~\ref{fig:S5}). The next dominant contribution to $\rho_{xx}$ comes from the $6\varphi$ component, which indicates the hexagonal symmetry of the basal plane. Moreover, a substantial $3\varphi$ contribution to $\rho_{xy}$ confirms the AHE originating from the altermagnetic order in MnTe. All of these results clearly indicate that crystal symmetry plays an important role in the in-plane angular magnetoresistance of MnTe films.

% fitted the curves with extract different components by performing a Fourier analysis by fitting the experimental data with a sum of $A_{\rm n}{\rm sin}({\rm n} \varphi + \varphi _{\rm n})$ for the transverse response and $A_{\rm n}{\rm cos}({\rm n} \varphi + \varphi _{\rm n})$ for the longitudinal response, where $A_{\rm n}$ and $\varphi _{\rm n}$ are the amplitude and phase of the n$^{\rm th}$ component, respectively. The transverse signal is dominated by n $=$ 2 component (see SI Fig.~\ref{fig:S5}a) and the longitudinal signal contains n $=$ 2, 4, 6 components (see SI Fig.~\ref{fig:S5}b). These results agree well with the predictions provided previously.~\cite{tschirner2023saturation} The $\rho_{xy}$ follows majorly ${\rm sin}({\rm 2} \varphi )$ trend which is because of magnetic anisotropy which is established to be two-fold dependent in the in-plane direction as measured previously.~\cite{orlova2025magnetization} The $\rho_{xx}$ has major contributions from ${\rm sin}({\rm 2} \varphi )$ and ${\rm sin}({\rm 6} \varphi )$ which originates from magnetic and crystalline anisotropy, respectively.

\begin{figure}[t!]
\centering
\includegraphics*[width=\textwidth]{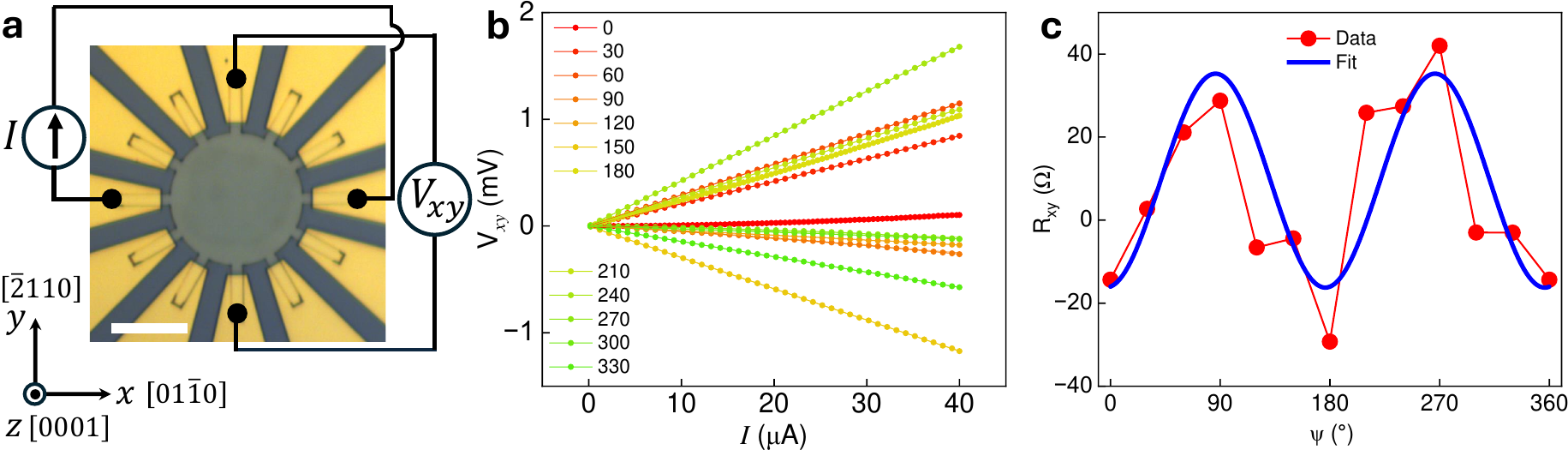}
\caption{\label{fig:4} \textbf{Crystallographic orientation dependence of transverse response.} \textbf{(a)} Image of the circular device and measurement geometry showing measurement at $\psi = 0$°. Here, $\psi$ is the angle between current and [01$\bar{1}$0] crystal direction. The measurement geometry is rotated in clockwise direction to get higher $\psi$ angles at 30° step. Scale bar is 20 µm. \textbf{(b)} Transverse voltage $V_{xy}$ measured as a function of the current flowing along different $\psi$ values. \textbf{(c)} Corresponding transverse resistance $R_{xy}$ as a function of $\psi$. All the measurements were performed at T = 175 K (below Néel temperature).} 
\end{figure}

Moreover, we have performed the angular dependence in the zx and yz planes by varying the angle $\theta$ and $\alpha$  with respect to the $z$-axis, respectively (Fig.~\ref{fig:3}d,g). This enables us to distinguish the contributions from crystal symmetry and purely magnetic effects. The $\rho_{xy}$ in both the zx and yz scans reveals a similar angular dependence with substantial magnetic anisotropy of the Néel vector [see Fig.~\ref{fig:3}e,h]. We observe a large signal for the fields along $z$-axis, which may be due to out-of-plane orbital magnetization in MnTe films.~\cite{ye2025dominantorbitalmagnetizationprototypical} In contrast, we observe distinct angular dependencies of $\rho_{xx}$ when rotating $\theta$ and $\alpha$. While sharp dips (Fig.~\ref{fig:3}f) and peaks (Fig.~\ref{fig:3}i) at 0°, 180°, and 360° are attributed to orbital magnetoresistance, additional smaller features appear at 90° and 270°, which corresponds to the field lying in-plane. Therefore, although the Néel vector rotates with the magnetic field, it remains influenced by the underlying crystal anisotropy. This is further supported by the fact that both the longitudinal and transverse responses exhibit weak angular dependence for magnetic fields (i.e., $B$ $\leq$ 1 T), which are insufficient to induce coherent rotation of the Néel vector.

% Both the zx and yz scns show principly a similar angular dependence revealing in-plane anisotropy of MnTe Néel vector. In the -1, 0 and 1 T fields, there was a very small change in longitudinal and transverse resistance suggesting that Néel vector saturates after 1 T field. In this regime, the anisotropy dominates and the applied field leads to small canting of the moments eventually leading to a small modulation in the longitudinal and transverse resistivity. At fields higher than 1 T, the field leads to spin-flop transition. In this regime, the Néel vector tends to rotate with the rotation in field and hence a much larger modulation in the resistivity.  However, the exact direction of the Néel vector remains influenced by the underlying crystal anisotropy, which is the reason for the higher order variations observed in the resistivity around 90$^{\circ}$ and 270$^{\circ}$ corresponding to the field being oriented in the in-plane direction.

To further understand the role of the Néel vector and crystal symmetry in electronic transport, we investigate the transverse response for current flowing along different crystallographic directions. We fabricate a circular disc device with 12 electrodes as depicted in Fig.~\ref{fig:4}a. The transverse voltage ($V_{xy}$) at zero magnetic field is measured while the current ($I$) is injected through one of the electrodes at angle $\psi$ with respect to [01$\bar{1}$0] crystal direction. We observe a linear current bias dependence of $V_{xy}$ [Fig.~\ref{fig:4}b]. The corresponding longitudinal resistance ($R_{xy}$) for different $\psi$ is presented in Fig.~\ref{fig:4}c, and can be well fitted with a function of the form $\sin(2\psi)$. Notably, this $\psi$ dependence mirrors the dominant $\sin(2\varphi)$ contribution observed in the magnetotransport measured with in-plane fields shown earlier in Fig.~\ref{fig:3}b. Unlike the measurement shown in Fig.~\ref{fig:3}a, the Néel vector remains fixed - presumably along a crystal axis, due to the absence of an external magnetic field. As a result, the transverse response depends on the relative orientation between the current and the Néel vector [Fig.~\ref{fig:4}a].

% the current direction is varied by changing the contacts in the circular device. Hence, the two cases are equivalent in which a relative angle between current and Néel vector is being varied and they show the same angle dependence. This clearly establishes that transverse response is dependent on the relative direction between current and the Néel vector [Fig.~\ref{fig:4}a]. It can act as a control on the transverse response and could be utilized for practical applications.

\subsection{Summary}

In summary, this work establishes the role of relative orientation of the crystal symmetry and altermagnetic order in determining the magneto-transport responses of epitaxial MnTe thin films. Both longitudinal ($\rho_{xx}$) and transverse ($\rho_{xy}$) responses are predominantly governed by a $2\varphi$ component, which reflects the magnetic anisotropy of the system. The next dominant contribution to $\rho_{xx}$ comes from the $6\varphi$ component, attributed to the crystalline anisotropy from the hexagonal symmetry of the basal plane. Moreover, a substantial $3\varphi$ contribution to $\rho_{xy}$ confirms the AHE originating from the altermagnetic order in MnTe. These results contribute significantly to the understanding of the fundamental magnetotransport characteristics of altermagnet MnTe and offer a basis for exploring its potential in future device architectures that exploit altermagnetic and crystal anisotropy-dependent transport.

% three response depends mainly on the Néel vector direction establishing its origin from Berry curvature. The longitudinal response depends on both the magnetic as well as crystallographic anisotropy. 

\subsection{Methods}

\textbf{Molecular beam epitaxy growth:} The MnTe layer was grown on InP(111) semi-insulating substrates in a Molecular Beam Epitaxy (MBE) chamber with base pressure  7$\times$10${^{-10}}$  mBar. Mn was evaporated from an e-gun with a rate of 0.1 $\mathring{\rm A}$/sec and Te was evaporated from a thermal cracker cell at a rate of 0.7 $\mathring{\rm A}$/sec. The InP wafer was prepared by thermal desorption under vacuum at 450 $^{\circ}$C until a x2 reconstruction appears in the RHEED indicative of a clean surface. The MnTe was grown at the same temperature of 450 $^{\circ}$C. The crystalline quality of the grown surface was monitored in-situ by RHEED. The top sample surface was in situ capped with a 2 nm Al$_2$O$_3$ layer to protect it from degradation with time.

\textbf{Device fabrication:} The devices were patterned into Hall-bar geometry using electron-beam lithography (EBL) and Ar ion milling and Ti/Au contacts were prepared by EBL and electron beam evaporation. 

\textbf{Transport measurements:} Magneto-transport properties were investigated using Quantum Design cryogen-free physical property measurement system (DynaCool) with an external electronic connection to Lockin SR830 to measure the longitudinal and transverse response simultaneously. We used a low frequency (213.3 Hz) a.c. current of 40 $\mu$A to measure the electronic transport.  

\subsection{Acknowledgements}

Authors acknowledge funding from European Innovation Council project 2DSPIN-TECH (No. 101135853), Graphene Flagship, 2D TECH VINNOVA competence center (No. 2019-00068), Wallenberg Initiative Materials Science for Sustainability (WISE) funded by the Knut and Alice Wallenberg (KAW) Foundation, Swedish Research Council (VR) grant (No. 2021–04821), FLAG-ERA project 2DSOTECH (VR No. 2021-05925) and MagicTune, Carl Tryggers foundation, Graphene Center, Areas of Advance (AoA) Nano, AoA Materials Science and AoA Energy programs at Chalmers University of Technology. The fabrication of devices was performed at the Nanofabrication laboratory MyFab at the Chalmers University of Technology. 

\subsection{Data availability}

The data that support the findings of this study are available from the corresponding authors on a reasonable request.

\subsection{Corresponding author}
Correspondence to Saroj P. Dash (saroj.dash@chalmers.se)

\subsection{Contributions}

H.B., A.D., S.P.D. conceived the idea and designed the experiments. H.B. fabricated and characterized the devices with L.P.'s help.  A.D., A.L., and P.T. did the material growth using MBE. A.K. performed the XRD measurements. H.B., P.K.R. and S.P.D. analyzed and interpreted the experimental data and wrote the manuscript with comments from all the authors. S.P.D. coordinated and supervised the project.

\subsection{Competing interests}
The authors declare no competing interests.

\bibliography{2_Main_bib}

\providecommand{\latin}[1]{#1}
\makeatletter
\providecommand{\doi}
  {\begingroup\let\do\@makeother\dospecials
  \catcode`\{=1 \catcode`\}=2 \doi@aux}
\providecommand{\doi@aux}[1]{\endgroup\texttt{#1}}
\makeatother
\providecommand*\mcitethebibliography{\thebibliography}
\csname @ifundefined\endcsname{endmcitethebibliography}  {\let\endmcitethebibliography\endthebibliography}{}
\begin{mcitethebibliography}{42}
\providecommand*\natexlab[1]{#1}
\providecommand*\mciteSetBstSublistMode[1]{}
\providecommand*\mciteSetBstMaxWidthForm[2]{}
\providecommand*\mciteBstWouldAddEndPuncttrue
  {\def\EndOfBibitem{\unskip.}}
\providecommand*\mciteBstWouldAddEndPunctfalse
  {\let\EndOfBibitem\relax}
\providecommand*\mciteSetBstMidEndSepPunct[3]{}
\providecommand*\mciteSetBstSublistLabelBeginEnd[3]{}
\providecommand*\EndOfBibitem{}
\mciteSetBstSublistMode{f}
\mciteSetBstMaxWidthForm{subitem}{(\alph{mcitesubitemcount})}
\mciteSetBstSublistLabelBeginEnd
  {\mcitemaxwidthsubitemform\space}
  {\relax}
  {\relax}

\bibitem[{\v{S}}mejkal \latin{et~al.}(2022){\v{S}}mejkal, Sinova, and Jungwirth]{vsmejkal2022beyond}
{\v{S}}mejkal,~L.; Sinova,~J.; Jungwirth,~T. Beyond conventional ferromagnetism and antiferromagnetism: A phase with nonrelativistic spin and crystal rotation symmetry. \emph{Phys. Rev. X} \textbf{2022}, \emph{12}, 031042\relax
\mciteBstWouldAddEndPuncttrue
\mciteSetBstMidEndSepPunct{\mcitedefaultmidpunct}
{\mcitedefaultendpunct}{\mcitedefaultseppunct}\relax
\EndOfBibitem
\bibitem[{\v{S}}mejkal \latin{et~al.}(2022){\v{S}}mejkal, Sinova, and Jungwirth]{vsmejkal2022emerging}
{\v{S}}mejkal,~L.; Sinova,~J.; Jungwirth,~T. Emerging research landscape of altermagnetism. \emph{Phys. Rev. X} \textbf{2022}, \emph{12}, 040501\relax
\mciteBstWouldAddEndPuncttrue
\mciteSetBstMidEndSepPunct{\mcitedefaultmidpunct}
{\mcitedefaultendpunct}{\mcitedefaultseppunct}\relax
\EndOfBibitem
\bibitem[Yuan \latin{et~al.}(2021)Yuan, Wang, Luo, and Zunger]{yuan2021prediction}
Yuan,~L.-D.; Wang,~Z.; Luo,~J.-W.; Zunger,~A. Prediction of low-Z collinear and noncollinear antiferromagnetic compounds having momentum-dependent spin splitting even without spin-orbit coupling. \emph{Phys. Rev. Mater.} \textbf{2021}, \emph{5}, 014409\relax
\mciteBstWouldAddEndPuncttrue
\mciteSetBstMidEndSepPunct{\mcitedefaultmidpunct}
{\mcitedefaultendpunct}{\mcitedefaultseppunct}\relax
\EndOfBibitem
\bibitem[Krempask{\`y} \latin{et~al.}(2024)Krempask{\`y}, {\v{S}}mejkal, D’souza, Hajlaoui, Springholz, Uhl{\'\i}{\v{r}}ov{\'a}, Alarab, Constantinou, Strocov, Usanov, \latin{et~al.} others]{krempasky2024altermagnetic}
Krempask{\`y},~J.; {\v{S}}mejkal,~L.; D’souza,~S.; Hajlaoui,~M.; Springholz,~G.; Uhl{\'\i}{\v{r}}ov{\'a},~K.; Alarab,~F.; Constantinou,~P.; Strocov,~V.; Usanov,~D.; others Altermagnetic lifting of Kramers spin degeneracy. \emph{Nature} \textbf{2024}, \emph{626}, 517--522\relax
\mciteBstWouldAddEndPuncttrue
\mciteSetBstMidEndSepPunct{\mcitedefaultmidpunct}
{\mcitedefaultendpunct}{\mcitedefaultseppunct}\relax
\EndOfBibitem
\bibitem[Mazin(2023)]{mazin2023altermagnetism}
Mazin,~I. Altermagnetism in MnTe: Origin, predicted manifestations, and routes to detwinning. \emph{Phys. Rev. B} \textbf{2023}, \emph{107}, L100418\relax
\mciteBstWouldAddEndPuncttrue
\mciteSetBstMidEndSepPunct{\mcitedefaultmidpunct}
{\mcitedefaultendpunct}{\mcitedefaultseppunct}\relax
\EndOfBibitem
\bibitem[Feng \latin{et~al.}(2022)Feng, Zhou, {\v{S}}mejkal, Wu, Zhu, Guo, Gonz{\'a}lez-Hern{\'a}ndez, Wang, Yan, Qin, \latin{et~al.} others]{feng2022anomalous}
Feng,~Z.; Zhou,~X.; {\v{S}}mejkal,~L.; Wu,~L.; Zhu,~Z.; Guo,~H.; Gonz{\'a}lez-Hern{\'a}ndez,~R.; Wang,~X.; Yan,~H.; Qin,~P.; others An anomalous Hall effect in altermagnetic ruthenium dioxide. \emph{Nat. Electron.} \textbf{2022}, \emph{5}, 735--743\relax
\mciteBstWouldAddEndPuncttrue
\mciteSetBstMidEndSepPunct{\mcitedefaultmidpunct}
{\mcitedefaultendpunct}{\mcitedefaultseppunct}\relax
\EndOfBibitem
\bibitem[{\v{S}}mejkal \latin{et~al.}(2020){\v{S}}mejkal, Gonz{\'a}lez-Hern{\'a}ndez, Jungwirth, and Sinova]{vsmejkal2020crystal}
{\v{S}}mejkal,~L.; Gonz{\'a}lez-Hern{\'a}ndez,~R.; Jungwirth,~T.; Sinova,~J. Crystal time-reversal symmetry breaking and spontaneous Hall effect in collinear antiferromagnets. \emph{Sci. Adv.} \textbf{2020}, \emph{6}, eaaz8809\relax
\mciteBstWouldAddEndPuncttrue
\mciteSetBstMidEndSepPunct{\mcitedefaultmidpunct}
{\mcitedefaultendpunct}{\mcitedefaultseppunct}\relax
\EndOfBibitem
\bibitem[Bai \latin{et~al.}(2024)Bai, Feng, Liu, {\v{S}}mejkal, Mokrousov, and Yao]{bai2024altermagnetism}
Bai,~L.; Feng,~W.; Liu,~S.; {\v{S}}mejkal,~L.; Mokrousov,~Y.; Yao,~Y. Altermagnetism: Exploring new frontiers in magnetism and spintronics. \emph{Adv. Funct. Mater.} \textbf{2024}, \emph{34}, 2409327\relax
\mciteBstWouldAddEndPuncttrue
\mciteSetBstMidEndSepPunct{\mcitedefaultmidpunct}
{\mcitedefaultendpunct}{\mcitedefaultseppunct}\relax
\EndOfBibitem
\bibitem[Guo \latin{et~al.}(2023)Guo, Liu, Janson, Fulga, van~den Brink, and Facio]{guo2023spin}
Guo,~Y.; Liu,~H.; Janson,~O.; Fulga,~I.~C.; van~den Brink,~J.; Facio,~J.~I. Spin-split collinear antiferromagnets: A large-scale ab-initio study. \emph{Mater. Today Phys.} \textbf{2023}, \emph{32}, 100991\relax
\mciteBstWouldAddEndPuncttrue
\mciteSetBstMidEndSepPunct{\mcitedefaultmidpunct}
{\mcitedefaultendpunct}{\mcitedefaultseppunct}\relax
\EndOfBibitem
\bibitem[Lee \latin{et~al.}(2024)Lee, Lee, Jung, Jung, Kim, Lee, Seok, Kim, Park, {\v{S}}mejkal, \latin{et~al.} others]{lee2024broken}
Lee,~S.; Lee,~S.; Jung,~S.; Jung,~J.; Kim,~D.; Lee,~Y.; Seok,~B.; Kim,~J.; Park,~B.~G.; {\v{S}}mejkal,~L.; others Broken Kramers degeneracy in altermagnetic MnTe. \emph{Phys. Rev. Lett.} \textbf{2024}, \emph{132}, 036702\relax
\mciteBstWouldAddEndPuncttrue
\mciteSetBstMidEndSepPunct{\mcitedefaultmidpunct}
{\mcitedefaultendpunct}{\mcitedefaultseppunct}\relax
\EndOfBibitem
\bibitem[Yang \latin{et~al.}(2025)Yang, Li, Yang, Li, Zheng, Zhu, Pan, Xu, Cao, Zhao, \latin{et~al.} others]{yang2025three}
Yang,~G.; Li,~Z.; Yang,~S.; Li,~J.; Zheng,~H.; Zhu,~W.; Pan,~Z.; Xu,~Y.; Cao,~S.; Zhao,~W.; others Three-dimensional mapping of the altermagnetic spin splitting in CrSb. \emph{Nat. Commun.} \textbf{2025}, \emph{16}, 1442\relax
\mciteBstWouldAddEndPuncttrue
\mciteSetBstMidEndSepPunct{\mcitedefaultmidpunct}
{\mcitedefaultendpunct}{\mcitedefaultseppunct}\relax
\EndOfBibitem
\bibitem[Reimers \latin{et~al.}(2024)Reimers, Odenbreit, {\v{S}}mejkal, Strocov, Constantinou, Hellenes, Jaeschke~Ubiergo, Campos, Bharadwaj, Chakraborty, \latin{et~al.} others]{reimers2024direct}
Reimers,~S.; Odenbreit,~L.; {\v{S}}mejkal,~L.; Strocov,~V.~N.; Constantinou,~P.; Hellenes,~A.~B.; Jaeschke~Ubiergo,~R.; Campos,~W.~H.; Bharadwaj,~V.~K.; Chakraborty,~A.; others Direct observation of altermagnetic band splitting in CrSb thin films. \emph{Nat. Commun.} \textbf{2024}, \emph{15}, 2116\relax
\mciteBstWouldAddEndPuncttrue
\mciteSetBstMidEndSepPunct{\mcitedefaultmidpunct}
{\mcitedefaultendpunct}{\mcitedefaultseppunct}\relax
\EndOfBibitem
\bibitem[Song \latin{et~al.}(2025)Song, Bai, Zhou, Han, Reichlova, Dil, Liu, Chen, and Pan]{song2025altermagnets}
Song,~C.; Bai,~H.; Zhou,~Z.; Han,~L.; Reichlova,~H.; Dil,~J.~H.; Liu,~J.; Chen,~X.; Pan,~F. Altermagnets as a new class of functional materials. \emph{Nat. Rev. Mater.} \textbf{2025}, 1--13\relax
\mciteBstWouldAddEndPuncttrue
\mciteSetBstMidEndSepPunct{\mcitedefaultmidpunct}
{\mcitedefaultendpunct}{\mcitedefaultseppunct}\relax
\EndOfBibitem
\bibitem[Cheong and Huang(2024)Cheong, and Huang]{cheong2024altermagnetism}
Cheong,~S.-W.; Huang,~F.-T. Altermagnetism with non-collinear spins. \emph{npj Quantum Mater.} \textbf{2024}, \emph{9}, 13\relax
\mciteBstWouldAddEndPuncttrue
\mciteSetBstMidEndSepPunct{\mcitedefaultmidpunct}
{\mcitedefaultendpunct}{\mcitedefaultseppunct}\relax
\EndOfBibitem
\bibitem[Sourounis and Manchon(2025)Sourounis, and Manchon]{sourounis2025efficient}
Sourounis,~K.; Manchon,~A. Efficient generation of spin currents in altermagnets via magnon drag. \emph{Phys. Rev. B} \textbf{2025}, \emph{111}, 134448\relax
\mciteBstWouldAddEndPuncttrue
\mciteSetBstMidEndSepPunct{\mcitedefaultmidpunct}
{\mcitedefaultendpunct}{\mcitedefaultseppunct}\relax
\EndOfBibitem
\bibitem[Bangar \latin{et~al.}(2023)Bangar, Khan, Kumar, Chowdhury, Muduli, and Muduli]{bangar2023large}
Bangar,~H.; Khan,~K. I.~A.; Kumar,~A.; Chowdhury,~N.; Muduli,~P.~K.; Muduli,~P.~K. Large spin hall conductivity in epitaxial thin films of kagome antiferromagnet Mn$_3$Sn at room temperature. \emph{Adv. Quantum Technol.} \textbf{2023}, \emph{6}, 2200115\relax
\mciteBstWouldAddEndPuncttrue
\mciteSetBstMidEndSepPunct{\mcitedefaultmidpunct}
{\mcitedefaultendpunct}{\mcitedefaultseppunct}\relax
\EndOfBibitem
\bibitem[Amin \latin{et~al.}(2024)Amin, Dal~Din, Golias, Niu, Zakharov, Fromage, Fields, Heywood, Cousins, Maccherozzi, \latin{et~al.} others]{amin2024nanoscale}
Amin,~O.; Dal~Din,~A.; Golias,~E.; Niu,~Y.; Zakharov,~A.; Fromage,~S.; Fields,~C.; Heywood,~S.; Cousins,~R.; Maccherozzi,~F.; others Nanoscale imaging and control of altermagnetism in MnTe. \emph{Nature} \textbf{2024}, \emph{636}, 348--353\relax
\mciteBstWouldAddEndPuncttrue
\mciteSetBstMidEndSepPunct{\mcitedefaultmidpunct}
{\mcitedefaultendpunct}{\mcitedefaultseppunct}\relax
\EndOfBibitem
\bibitem[Li \latin{et~al.}(2024)Li, Hu, Li, Wang, Chen, Thiagarajan, Leandersson, Polley, Kim, Liu, \latin{et~al.} others]{li2024topological}
Li,~C.; Hu,~M.; Li,~Z.; Wang,~Y.; Chen,~W.; Thiagarajan,~B.; Leandersson,~M.; Polley,~C.; Kim,~T.; Liu,~H.; others Topological weyl altermagnetism in CrSb. \emph{arXiv preprint arXiv:2405.14777} \textbf{2024}, \relax
\mciteBstWouldAddEndPunctfalse
\mciteSetBstMidEndSepPunct{\mcitedefaultmidpunct}
{}{\mcitedefaultseppunct}\relax
\EndOfBibitem
\bibitem[Hu \latin{et~al.}(2024)Hu, Janson, Felser, McClarty, Brink, and Vergniory]{hu2024spin}
Hu,~M.; Janson,~O.; Felser,~C.; McClarty,~P.; Brink,~J. v.~d.; Vergniory,~M.~G. Spin Hall and Edelstein Effects in Novel Chiral Noncollinear Altermagnets. \emph{arXiv preprint arXiv:2410.17993} \textbf{2024}, \relax
\mciteBstWouldAddEndPunctfalse
\mciteSetBstMidEndSepPunct{\mcitedefaultmidpunct}
{}{\mcitedefaultseppunct}\relax
\EndOfBibitem
\bibitem[Gomonay \latin{et~al.}(2024)Gomonay, Kravchuk, Jaeschke-Ubiergo, Yershov, Jungwirth, {\v{S}}mejkal, Brink, and Sinova]{gomonay2024structure}
Gomonay,~O.; Kravchuk,~V.; Jaeschke-Ubiergo,~R.; Yershov,~K.; Jungwirth,~T.; {\v{S}}mejkal,~L.; Brink,~J. v.~d.; Sinova,~J. Structure, control, and dynamics of altermagnetic textures. \emph{npj Spintronics} \textbf{2024}, \emph{2}, 35\relax
\mciteBstWouldAddEndPuncttrue
\mciteSetBstMidEndSepPunct{\mcitedefaultmidpunct}
{\mcitedefaultendpunct}{\mcitedefaultseppunct}\relax
\EndOfBibitem
\bibitem[Kravchuk \latin{et~al.}(2025)Kravchuk, Yershov, Facio, Guo, Janson, Gomonay, Sinova, and Brink]{kravchuk2025chiral}
Kravchuk,~V.~P.; Yershov,~K.~V.; Facio,~J.~I.; Guo,~Y.; Janson,~O.; Gomonay,~O.; Sinova,~J.; Brink,~J. v.~d. Chiral magnetic excitations and domain textures of $ g $-wave altermagnets. \emph{arXiv preprint arXiv:2504.05241} \textbf{2025}, \relax
\mciteBstWouldAddEndPunctfalse
\mciteSetBstMidEndSepPunct{\mcitedefaultmidpunct}
{}{\mcitedefaultseppunct}\relax
\EndOfBibitem
\bibitem[Kriegner \latin{et~al.}(2016)Kriegner, V{\`y}born{\`y}, Olejn{\'\i}k, Reichlov{\'a}, Nov{\'a}k, Marti, Gazquez, Saidl, N{\v{e}}mec, Volobuev, \latin{et~al.} others]{kriegner2016multiple}
Kriegner,~D.; V{\`y}born{\`y},~K.; Olejn{\'\i}k,~K.; Reichlov{\'a},~H.; Nov{\'a}k,~V.; Marti,~X.; Gazquez,~J.; Saidl,~V.; N{\v{e}}mec,~P.; Volobuev,~V.; others Multiple-stable anisotropic magnetoresistance memory in antiferromagnetic MnTe. \emph{Nat. Commun.} \textbf{2016}, \emph{7}, 11623\relax
\mciteBstWouldAddEndPuncttrue
\mciteSetBstMidEndSepPunct{\mcitedefaultmidpunct}
{\mcitedefaultendpunct}{\mcitedefaultseppunct}\relax
\EndOfBibitem
\bibitem[Gonzalez~Betancourt \latin{et~al.}(2023)Gonzalez~Betancourt, Zub{\'a}{\v{c}}, Gonzalez-Hernandez, Geishendorf, {\v{S}}ob{\'a}{\v{n}}, Springholz, Olejn{\'\i}k, {\v{S}}mejkal, Sinova, Jungwirth, \latin{et~al.} others]{gonzalez2023spontaneous}
Gonzalez~Betancourt,~R.; Zub{\'a}{\v{c}},~J.; Gonzalez-Hernandez,~R.; Geishendorf,~K.; {\v{S}}ob{\'a}{\v{n}},~Z.; Springholz,~G.; Olejn{\'\i}k,~K.; {\v{S}}mejkal,~L.; Sinova,~J.; Jungwirth,~T.; others Spontaneous anomalous Hall effect arising from an unconventional compensated magnetic phase in a semiconductor. \emph{Phys. Rev. Lett.} \textbf{2023}, \emph{130}, 036702\relax
\mciteBstWouldAddEndPuncttrue
\mciteSetBstMidEndSepPunct{\mcitedefaultmidpunct}
{\mcitedefaultendpunct}{\mcitedefaultseppunct}\relax
\EndOfBibitem
\bibitem[Osumi \latin{et~al.}(2024)Osumi, Souma, Aoyama, Yamauchi, Honma, Nakayama, Takahashi, Ohgushi, and Sato]{osumi2024observation}
Osumi,~T.; Souma,~S.; Aoyama,~T.; Yamauchi,~K.; Honma,~A.; Nakayama,~K.; Takahashi,~T.; Ohgushi,~K.; Sato,~T. Observation of a giant band splitting in altermagnetic MnTe. \emph{Phys. Rev. B} \textbf{2024}, \emph{109}, 115102\relax
\mciteBstWouldAddEndPuncttrue
\mciteSetBstMidEndSepPunct{\mcitedefaultmidpunct}
{\mcitedefaultendpunct}{\mcitedefaultseppunct}\relax
\EndOfBibitem
\bibitem[Lovesey \latin{et~al.}(2023)Lovesey, Khalyavin, and Van Der~Laan]{lovesey2023templates}
Lovesey,~S.; Khalyavin,~D.; Van Der~Laan,~G. Templates for magnetic symmetry and altermagnetism in hexagonal MnTe. \emph{Phys. Rev. B} \textbf{2023}, \emph{108}, 174437\relax
\mciteBstWouldAddEndPuncttrue
\mciteSetBstMidEndSepPunct{\mcitedefaultmidpunct}
{\mcitedefaultendpunct}{\mcitedefaultseppunct}\relax
\EndOfBibitem
\bibitem[Takahashi \latin{et~al.}(2025)Takahashi, Huang, Yu, and Zang]{takahashi2025symmetry}
Takahashi,~K.; Huang,~H.-F.; Yu,~J.-X.; Zang,~J. Symmetry and Minimal Hamiltonian of Nonsymmorphic Collinear Antiferromagnet MnTe. \emph{arXiv preprint arXiv:2503.07951} \textbf{2025}, \relax
\mciteBstWouldAddEndPunctfalse
\mciteSetBstMidEndSepPunct{\mcitedefaultmidpunct}
{}{\mcitedefaultseppunct}\relax
\EndOfBibitem
\bibitem[Kluczyk \latin{et~al.}(2024)Kluczyk, Gas, Grzybowski, Skupi{\'n}ski, Borysiewicz, F{\k{a}}s, Suffczy{\'n}ski, Domagala, Grasza, Mycielski, \latin{et~al.} others]{kluczyk2024coexistence}
Kluczyk,~K.; Gas,~K.; Grzybowski,~M.; Skupi{\'n}ski,~P.; Borysiewicz,~M.; F{\k{a}}s,~T.; Suffczy{\'n}ski,~J.; Domagala,~J.; Grasza,~K.; Mycielski,~A.; others Coexistence of anomalous Hall effect and weak magnetization in a nominally collinear antiferromagnet MnTe. \emph{Phys. Rev. B} \textbf{2024}, \emph{110}, 155201\relax
\mciteBstWouldAddEndPuncttrue
\mciteSetBstMidEndSepPunct{\mcitedefaultmidpunct}
{\mcitedefaultendpunct}{\mcitedefaultseppunct}\relax
\EndOfBibitem
\bibitem[Sato \latin{et~al.}(2024)Sato, Haddad, Fulga, Assaad, and van~den Brink]{sato2024altermagnetic}
Sato,~T.; Haddad,~S.; Fulga,~I.~C.; Assaad,~F.~F.; van~den Brink,~J. Altermagnetic anomalous Hall effect emerging from electronic correlations. \emph{Phys. Rev. Lett.} \textbf{2024}, \emph{133}, 086503\relax
\mciteBstWouldAddEndPuncttrue
\mciteSetBstMidEndSepPunct{\mcitedefaultmidpunct}
{\mcitedefaultendpunct}{\mcitedefaultseppunct}\relax
\EndOfBibitem
\bibitem[Xiao \latin{et~al.}(2010)Xiao, Chang, and Niu]{xiao2010berry}
Xiao,~D.; Chang,~M.-C.; Niu,~Q. Berry phase effects on electronic properties. \emph{Rev. Mod. Phys.} \textbf{2010}, \emph{82}, 1959--2007\relax
\mciteBstWouldAddEndPuncttrue
\mciteSetBstMidEndSepPunct{\mcitedefaultmidpunct}
{\mcitedefaultendpunct}{\mcitedefaultseppunct}\relax
\EndOfBibitem
\bibitem[{\v{S}}mejkal \latin{et~al.}(2022){\v{S}}mejkal, MacDonald, Sinova, Nakatsuji, and Jungwirth]{vsmejkal2022anomalous}
{\v{S}}mejkal,~L.; MacDonald,~A.~H.; Sinova,~J.; Nakatsuji,~S.; Jungwirth,~T. Anomalous hall antiferromagnets. \emph{Nat. Rev. Mater.} \textbf{2022}, \emph{7}, 482--496\relax
\mciteBstWouldAddEndPuncttrue
\mciteSetBstMidEndSepPunct{\mcitedefaultmidpunct}
{\mcitedefaultendpunct}{\mcitedefaultseppunct}\relax
\EndOfBibitem
\bibitem[Tschirner \latin{et~al.}(2023)Tschirner, Ke{\ss}ler, Gonzalez~Betancourt, Kotte, Kriegner, B{\"u}chner, Dufouleur, Kamp, Jovic, Smejkal, \latin{et~al.} others]{tschirner2023saturation}
Tschirner,~T.; Ke{\ss}ler,~P.; Gonzalez~Betancourt,~R.~D.; Kotte,~T.; Kriegner,~D.; B{\"u}chner,~B.; Dufouleur,~J.; Kamp,~M.; Jovic,~V.; Smejkal,~L.; others Saturation of the anomalous Hall effect at high magnetic fields in altermagnetic RuO$_2$. \emph{APL Mater.} \textbf{2023}, \emph{11}\relax
\mciteBstWouldAddEndPuncttrue
\mciteSetBstMidEndSepPunct{\mcitedefaultmidpunct}
{\mcitedefaultendpunct}{\mcitedefaultseppunct}\relax
\EndOfBibitem
\bibitem[Vil()]{Villars2023:sm_isp_sd_0530446}
MnTe Crystal Structure: Datasheet from ``PAULING FILE Multinaries Edition -- 2022'' in SpringerMaterials (https://materials.springer.com/isp/crystallographic/docs/sd{\_}0530446). \url{https://materials.springer.com/isp/crystallographic/docs/sd_0530446}\relax
\mciteBstWouldAddEndPuncttrue
\mciteSetBstMidEndSepPunct{\mcitedefaultmidpunct}
{\mcitedefaultendpunct}{\mcitedefaultseppunct}\relax
\EndOfBibitem
\bibitem[Vil()]{Villars2023:sm_isp_sd_0453882}
InP Crystal Structure: Datasheet from ``PAULING FILE Multinaries Edition -- 2022'' in SpringerMaterials (https://materials.springer.com/isp/crystallographic/docs/sd{\_}0453882). \url{https://materials.springer.com/isp/crystallographic/docs/sd_0453882}\relax
\mciteBstWouldAddEndPuncttrue
\mciteSetBstMidEndSepPunct{\mcitedefaultmidpunct}
{\mcitedefaultendpunct}{\mcitedefaultseppunct}\relax
\EndOfBibitem
\bibitem[Magnin and Diep(2012)Magnin, and Diep]{magnin2012monte}
Magnin,~Y.; Diep,~H.~T. Monte Carlo study of magnetic resistivity in semiconducting MnTe. \emph{Phys. Rev. B} \textbf{2012}, \emph{85}, 184413\relax
\mciteBstWouldAddEndPuncttrue
\mciteSetBstMidEndSepPunct{\mcitedefaultmidpunct}
{\mcitedefaultendpunct}{\mcitedefaultseppunct}\relax
\EndOfBibitem
\bibitem[Kriegner \latin{et~al.}(2017)Kriegner, Reichlova, Grenzer, Schmidt, Ressouche, Godinho, Wagner, Martin, Shick, Volobuev, \latin{et~al.} others]{kriegner2017magnetic}
Kriegner,~D.; Reichlova,~H.; Grenzer,~J.; Schmidt,~W.; Ressouche,~E.; Godinho,~J.; Wagner,~T.; Martin,~S.; Shick,~A.; Volobuev,~V.; others Magnetic anisotropy in antiferromagnetic hexagonal MnTe. \emph{Phys. Rev. B} \textbf{2017}, \emph{96}, 214418\relax
\mciteBstWouldAddEndPuncttrue
\mciteSetBstMidEndSepPunct{\mcitedefaultmidpunct}
{\mcitedefaultendpunct}{\mcitedefaultseppunct}\relax
\EndOfBibitem
\bibitem[Bogdanov \latin{et~al.}(2007)Bogdanov, Zhuravlev, and R{\"o}{\ss}ler]{bogdanov2007spin}
Bogdanov,~A.; Zhuravlev,~A.; R{\"o}{\ss}ler,~U. Spin-flop transition in uniaxial antiferromagnets: Magnetic phases, reorientation effects, and multidomain states. \emph{Phys. Rev. B} \textbf{2007}, \emph{75}, 094425\relax
\mciteBstWouldAddEndPuncttrue
\mciteSetBstMidEndSepPunct{\mcitedefaultmidpunct}
{\mcitedefaultendpunct}{\mcitedefaultseppunct}\relax
\EndOfBibitem
\bibitem[Rout \latin{et~al.}(2017)Rout, Agireen, Maniv, Goldstein, and Dagan]{rout2017sixfold}
Rout,~P.~K.; Agireen,~I.; Maniv,~E.; Goldstein,~M.; Dagan,~Y. Six-fold crystalline anisotropic magnetoresistance in the (111) ${\mathbf{LaAlO}}_{3}/{\mathbf{SrTiO}}_{3}$ oxide interface. \emph{Phys. Rev. B} \textbf{2017}, \emph{95}, 241107\relax
\mciteBstWouldAddEndPuncttrue
\mciteSetBstMidEndSepPunct{\mcitedefaultmidpunct}
{\mcitedefaultendpunct}{\mcitedefaultseppunct}\relax
\EndOfBibitem
\bibitem[Gonzalez~Betancourt \latin{et~al.}(2024)Gonzalez~Betancourt, Zub{\'a}{\v{c}}, Geishendorf, Ritzinger, R{\u u}{\v{z}}i{\v{c}}kov{\'a}, Kotte, {\v{Z}}elezn{\`y}, Olejn{\'\i}k, Springholz, B{\"u}chner, \latin{et~al.} others]{gonzalez2024anisotropic}
Gonzalez~Betancourt,~R.~D.; Zub{\'a}{\v{c}},~J.; Geishendorf,~K.; Ritzinger,~P.; R{\u u}{\v{z}}i{\v{c}}kov{\'a},~B.; Kotte,~T.; {\v{Z}}elezn{\`y},~J.; Olejn{\'\i}k,~K.; Springholz,~G.; B{\"u}chner,~B.; others Anisotropic magnetoresistance in altermagnetic MnTe. \emph{npj Spintronics} \textbf{2024}, \emph{2}, 45\relax
\mciteBstWouldAddEndPuncttrue
\mciteSetBstMidEndSepPunct{\mcitedefaultmidpunct}
{\mcitedefaultendpunct}{\mcitedefaultseppunct}\relax
\EndOfBibitem
\bibitem[Ritzinger and V{\`y}born{\`y}(2023)Ritzinger, and V{\`y}born{\`y}]{ritzinger2023anisotropic}
Ritzinger,~P.; V{\`y}born{\`y},~K. Anisotropic magnetoresistance: materials, models and applications. \emph{R. Soc. Open Sci.} \textbf{2023}, \emph{10}, 230564\relax
\mciteBstWouldAddEndPuncttrue
\mciteSetBstMidEndSepPunct{\mcitedefaultmidpunct}
{\mcitedefaultendpunct}{\mcitedefaultseppunct}\relax
\EndOfBibitem
\bibitem[McGuire and Potter(1975)McGuire, and Potter]{McGuire1975anisotropic}
McGuire,~T.; Potter,~R. Anisotropic magnetoresistance in ferromagnetic 3d alloys. \emph{IEEE Trans. Magn.} \textbf{1975}, \emph{11}, 1018--1038\relax
\mciteBstWouldAddEndPuncttrue
\mciteSetBstMidEndSepPunct{\mcitedefaultmidpunct}
{\mcitedefaultendpunct}{\mcitedefaultseppunct}\relax
\EndOfBibitem
\bibitem[Ye \latin{et~al.}(2025)Ye, Tenzin, Sławińska, and Autieri]{ye2025dominantorbitalmagnetizationprototypical}
Ye,~C.~C.; Tenzin,~K.; Sławińska,~J.; Autieri,~C. Dominant orbital magnetization in the prototypical altermagnet MnTe. 2025; \url{https://arxiv.org/abs/2505.08675}\relax
\mciteBstWouldAddEndPuncttrue
\mciteSetBstMidEndSepPunct{\mcitedefaultmidpunct}
{\mcitedefaultendpunct}{\mcitedefaultseppunct}\relax
\EndOfBibitem
\end{mcitethebibliography}

\begin{suppinfo}

\section{S1: X-ray diffraction}

\renewcommand{\thefigure}{S1}
\begin{figure}[H]
\centering
\includegraphics*[width=0.5\textwidth]{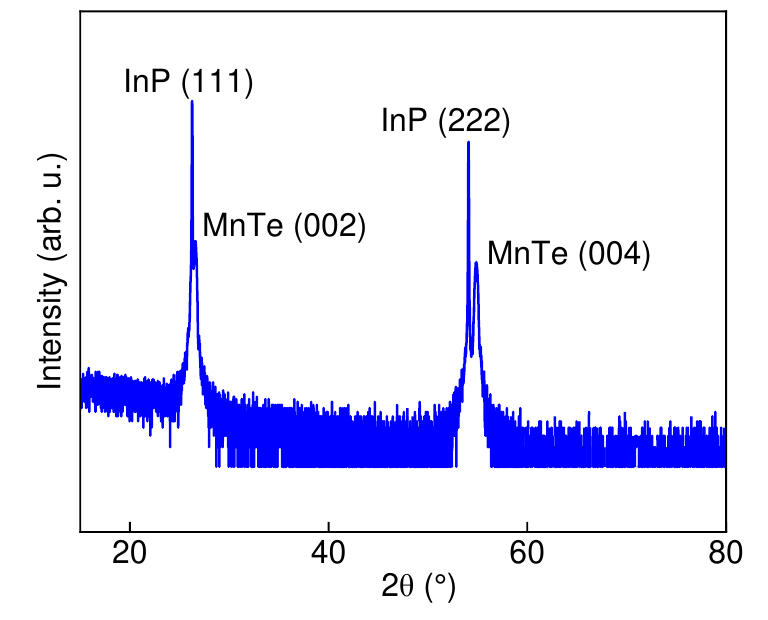}
\caption{\label{fig:S1} 2$\theta-\omega$ x-ray diffraction scan showing no extra peaks other than the MnTe and substrate InP. %\textbf{(b)} X-ray reflectivity data used to determine the film thickness and roughness.
} 
\end{figure}

We performed the full 2$\theta-\omega$ x-ray diffraction (XRD) scan as presented in Fig.~\ref{fig:S1}. The XRD data clearly show peaks only from MnTe (0002), (0004) and InP (111), (222) revealing good quality of our sample without any undesired peaks.

%Additionaly, we did x-ray reflectivity (XRR) scan to determine the film thickness and roughness. The data was fitted using aaaa model and the roughness and thickness are found to be xx and yy, respectively.

\section{S2: Magneto-resistance and anomalous Hall effect}

%\renewcommand{\thefigure}{S2}
%\begin{figure}[H]
%\centering
%\includegraphics*[width=0.7\textwidth]{FigS2_v2.pdf}
%\caption{\label{fig:S2} Magneto-resistance data at \textbf{(a)} 2 %K, \textbf{(b)} 25 K, \textbf{(c)} 75 K, and \textbf{(d)} 300 K.} 
%\end{figure}

%We presented magneto-resistance (MR) data at temperatures around T$_{\rm N}$ in the main manuscript Fig.~\ref{fig:2}. Here in Fig.~\ref{fig:S2}, we show MR at 2 K, 25 K, 75 K, and 300 K. At 300 K, the MR trend is typical to a paramagnetic phase. However, at temperatures around 100 K and below shows diverging curves probably due to enhanced anisotropy as explained in the main text.

%\renewcommand{\thefigure}{S3}
%\begin{figure}[H]
%\centering
%\includegraphics*[width=0.7\textwidth]{FigS3_v2.pdf}
%\caption{\label{fig:S3} Anomalous Hall effect (AHE) data at %\textbf{(a)} 75 K, \textbf{(b)}, and \textbf{(b)} 300 K.} 
%\end{figure}

The isolate the odd contribution to the anomalous Hall effect (AHE) signal, we first subtracted the even-in-field or symmetric component using the following formula:

\vspace{-3mm}
\begin{equation}\label{EqS1}
\begin{split}
\rho_{xy}^{\rm odd} = \frac{\rho (H) + \rho (-H)}{2},
\end{split}
\end{equation}

where, $\rho (H)$ and $\rho (-H)$ is AHE signal measured for forward sweep (-ve field to +ve field) and reverse sweep, respectively. Subsequently, we subtracted the ordinary Hall contribution by using the linear background at high field. %The AHE for 75 K and 300 K is presented in Fig.~\ref{fig:S3}a and b respectively.

The trend of saturated and spontaneous AHE is presented in Fig.~\ref{fig:S3}a and b, respectively. The trend is quite complex to explain. Above T$_{\rm N}$, the system is paramagnet and easy to saturate all the moments along the field hence a high saturated value but a negligible spontaneous AHE. Below T$_{\rm N}$, the altermagnet phase starts which leads to enhanced spontaneous AHE but a reduced saturated AHE. This is possible due to difficulty in saturating the moments towards the applied field due to altermganetic phase. Further variations below T$_{\rm N}$ could possibly origin from changes in resistivity and Fermi level with temperature.

\renewcommand{\thefigure}{S2}
\begin{figure}[H]
\centering
\includegraphics*[width=0.8\textwidth]{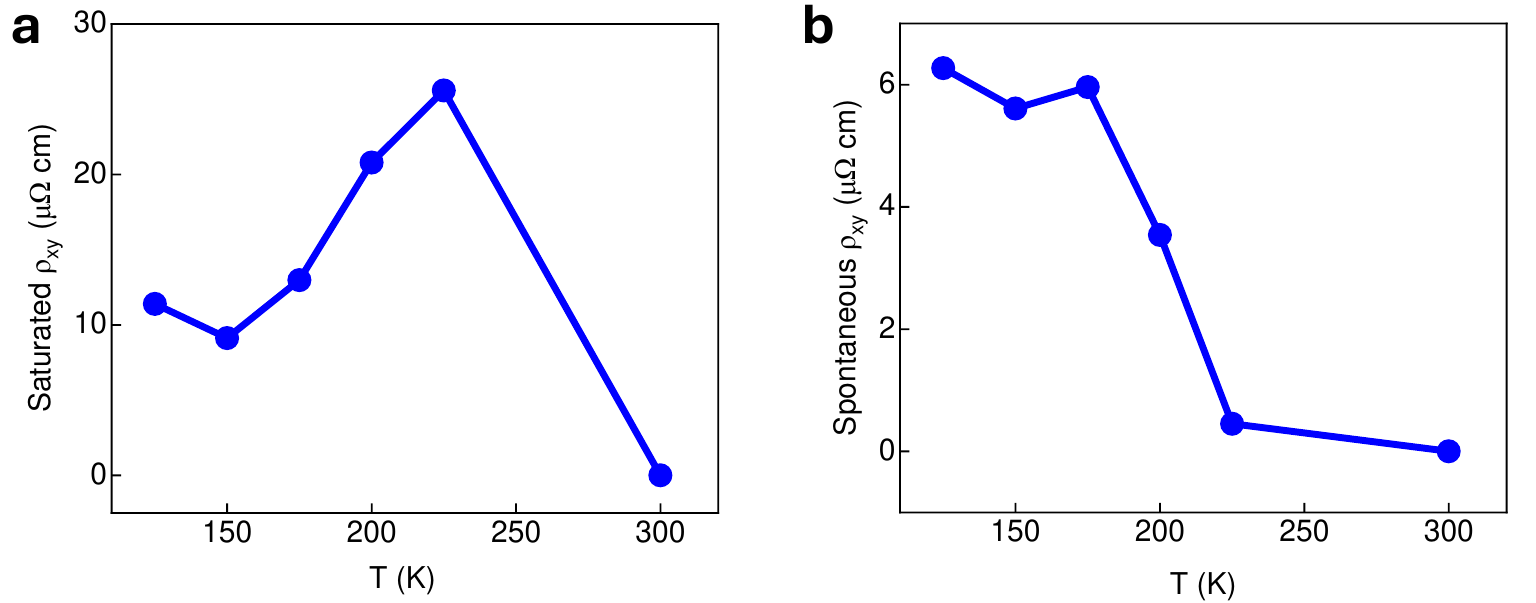}
\caption{\label{fig:S3} Temperature dependent \textbf{(a)} saturated and \textbf{(b)} spontaneous AHE resistivity.} 
\end{figure}

\renewcommand{\thefigure}{S3}
\begin{figure}[H]
\centering
\includegraphics*[width=1\textwidth]{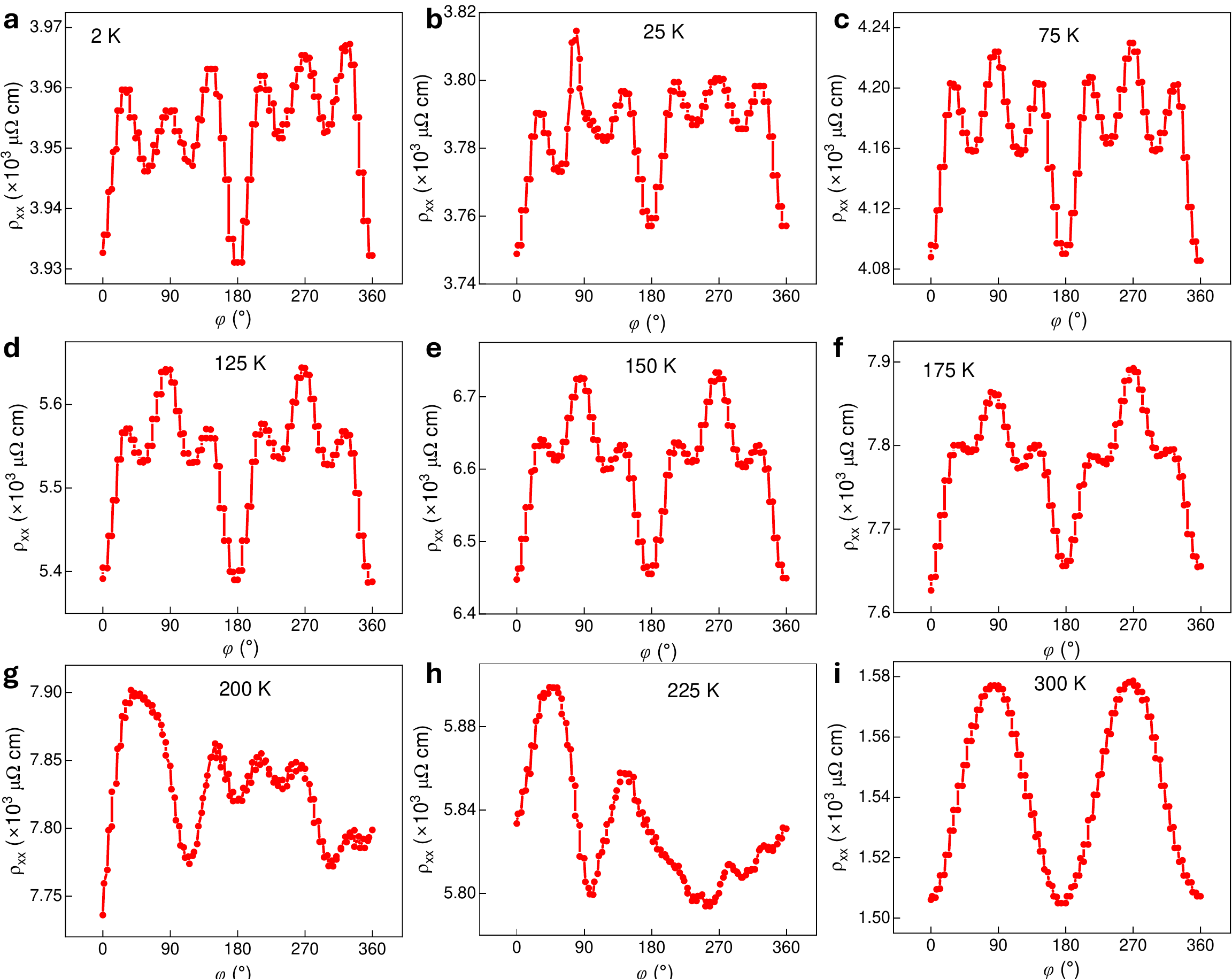}
\caption{\label{fig:S4} Anisotropic magneto-resistance (AMR) data at \textbf{(a)} 2 K, \textbf{(b)} 25 K, \textbf{(c)} 75 K, \textbf{(d)} 125 K, \textbf{(e)} 150 K, \textbf{(f)} 175 K, \textbf{(g)} 200 K, \textbf{(h)} 225 K, and \textbf{(i)} 300 K. AMR was performed with a 13 T field.} 
\end{figure}

The anisotropic magnetoresistance (AMR) response is presented in Fig.~\ref{fig:S4} at different temperatures with a constant 13 T field. AMR at several magnetic field is shown in Fig.~\ref{fig:3}c at 175 K. At temperatures, higher than T$_{_{\rm N}}$, the AMR follows two-fold dependence similar to non-crystalline AMR. Below T$_{_{\rm N}}$, the higher order terms start appearing from the crystalline AMR as well. The field dependence of AHE and AMR at 175 K is presented in Fig.~\ref{fig:3}b and c, respectively. The summary of the fitting results is presented in Fig.~\ref{fig:S5}. The transverse resistivity contains mainly two-fold dependence and it is nearly constant with field owing to magnetic anisotropy. In contrast, longitudinal resistivity contains two-fold, four-fold, and six-fold dependence owing to both magnetic and crystalline anisotropy.

\renewcommand{\thefigure}{S4}
\begin{figure}[H]
\centering
\includegraphics*[width=0.7\textwidth]{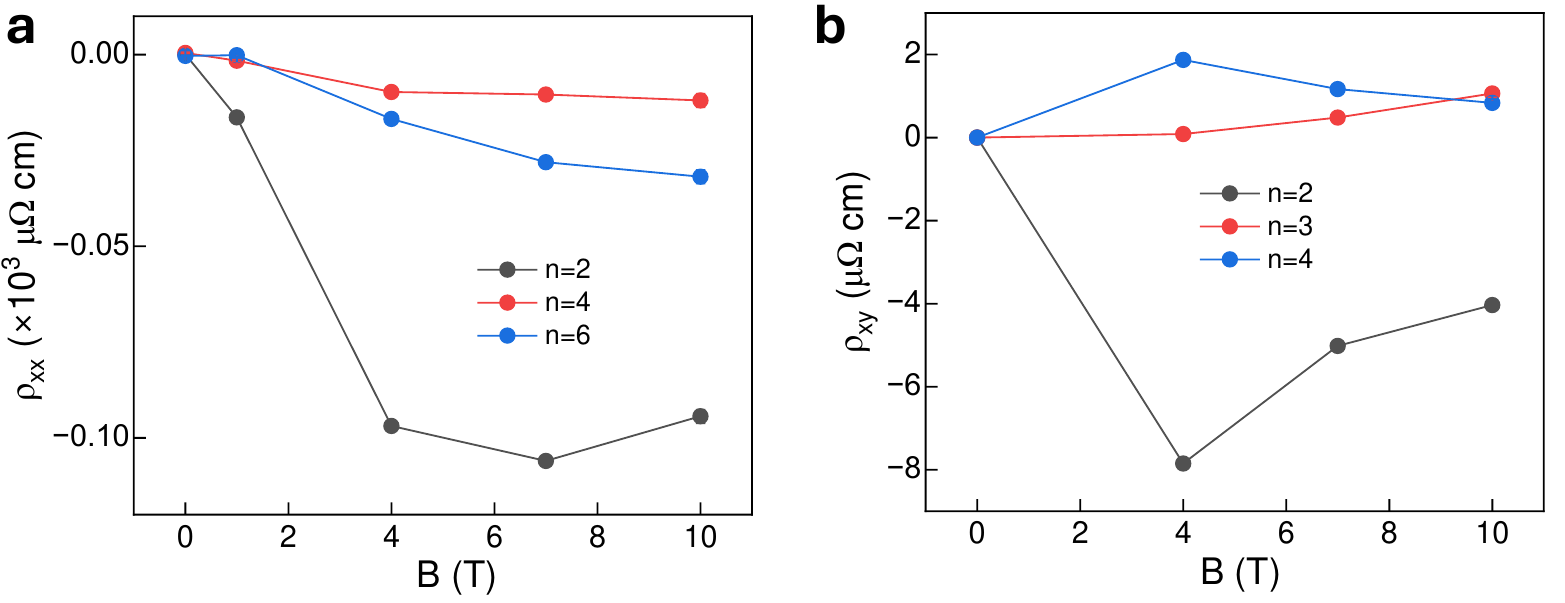}
\caption{\label{fig:S5} Magnetic field dependence of different components for \textbf{(a)} transverse and \textbf{(b)} longitudinal resistivites.} 
\end{figure}

\end{suppinfo}


\end{document}